%% file: tripathy.tex
\documentclass{svmult}
\usepackage{epsfig,graphicx} 
\usepackage[bottom]{footmisc}
\usepackage{natbib,url}      
\bibpunct{(}{)}{;}{a}{}{,}   
\def\cite#1{\citealp{#1}}    
\def\authorindex#1{}
\def\figspath{.}  

\begin{document}\newcount\preprintheader\preprintheader=1


\input{rr-latexdefs}  

\title*{Do Active Regions Modify Oscillation Frequencies?}


\author{S. C. Tripathy \and K. Jain \and F. Hill}

\authorindex{Tripathy, S. C.}
\authorindex{Jain, K.}
\authorindex{Hill, F.}


\institute{GONG Program, National Solar Observatory, Tucson, USA}

\maketitle

\setcounter{footnote}{0}  

\begin{abstract} 
 We investigate the variation of high-degree mode frequencies as a local
response to the active regions in two different phases of the solar activity
cycle. We find that the correlation between frequency shifts and the surface
magnetic activity  measured locally  are significantly different during the
two activity periods. 
\end{abstract}

\section{Introduction}      \label{tripathy-sec:introduction}

The oscillation frequencies are known to vary in phase with the solar
activity cycle.  In most of the earlier studies \citep[see][ and
references therein]{tripathy-2003SoPh..213..257J}, the variation
between the frequency shifts and activity, as measured by different
proxies, demonstrated a linear relation. But a detailed analysis using
the improved and continuously measured eigen-frequencies over solar
cycle 23 indicates complex relationships: a strong correlation during
the rising and declining phases and a significantly lower correlation
during the minimum phase \citep{tripathy-jain09}.  In addition, there
is no consensus as to the solar origin of these changes. There is some
indication that the variation of the high-degree mode frequencies are
spatially as well as temporally associated with active regions
\citep{tripathy-2000SoPh..192..363H, tripathy-2008ASPC..383..305H}.  In this
context, we investigate the
variation of high-degree mode frequencies as a local response to active
regions. 

\begin{figure}  
  \centering
  \includegraphics[width=10cm]{\figspath/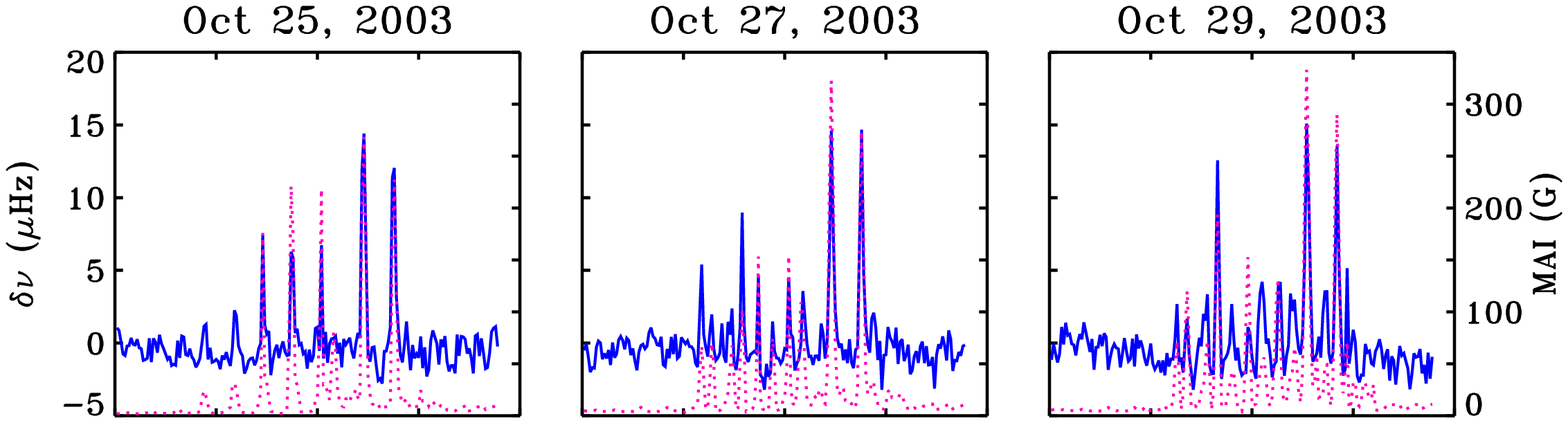}
\includegraphics[width=10cm]{\figspath/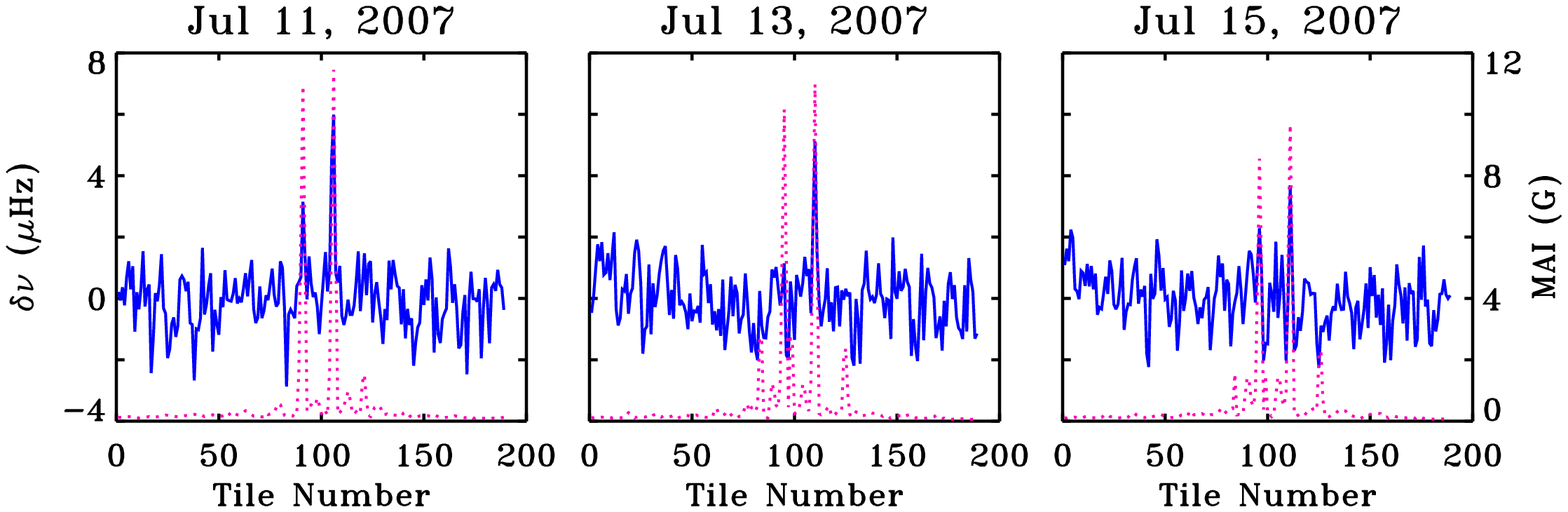}
   \caption[]{\label{tripathy-fig1}
 The frequency shifts  (solid line) for a single multiplet ($n$ =
3, $\ell$ = 242)  for 189 dense-pack regions. The upper and lower panels
represent CR 2009 and CR 2058, respectively. The dotted line denotes the MAI
and is calculated using MDI 96-minute magnetograms for CR 2009 and GONG
1-minute magnetograms for CR 2058.  The date on the top of the figures
indicates the beginning date of the ring-day analysis.  
}\end{figure}

\section{Analysis and Results}
We use the ring-diagram technique \citep{tripathy-1988ApJ...333..996H,
tripathy-2003ESASP.517..255C} to
calculate the high-degree mode frequencies and examine the frequency shifts 
during two Carrington Rotation (CR) periods of the solar cycle 23:
one during the descending phase  covering the period October
21--November 19, 2003 (CR 2009) and the other during the low activity period of 
June 20--July 18, 2007 (CR 2058). We analyze a set of 189 individual regions
on the solar disk, which is commonly referred to as a dense-pack mosaic. Each
region is about 15\deg\ $\times$ 15\deg\ in heliographic latitude and
longitude and is tracked for
a period of 1664 minutes.
The centers of the regions are separated by 7.5$\deg$\ in latitude and longitude
and extend roughly  to 52.5\deg\ from disk center.  By repeating the analysis
over the entire dense-pack mosaic, the frequencies are determined as a
function of ring-day (1664 min) and position on the solar disk. For each 
wavenumber and mode order, the frequency shifts ($\delta \nu$) are 
computed relative to the  spatial average obtained from the 189 tiles.  The
strength of the magnetic field associated with each tile is also 
estimated by calculating a Magnetic Activity Index (MAI) which represents the
average over all the pixels in a given tile. This is obtained from the 
magnetograms mapped and tracked in the same way as the Dopplergrams. 

 The frequency shifts  corresponding
to the $p_3$  ($n$ = 3) mode with harmonic degree $\ell$ = 242 are shown in
Figure~\ref{tripathy-fig1}. It reveals that the frequencies can vary as much as
20 $\mu$Hz within an active region having high magnetic-field strength
compared to the quiet Sun. In the same figure, we
also plot   the corresponding MAI (dotted line). A good agreement between
$\delta\nu$ and MAI, particularly for large shifts, is clearly visible. 

\begin{figure}  
  \centering
     \includegraphics[width=10cm]{\figspath/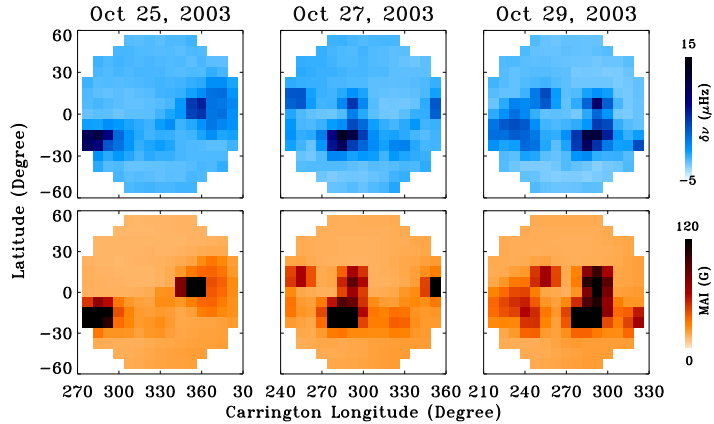}
    \caption[]{\label{tripathy-fig3}
 Average frequency shifts of 189 dense pack tiles (top
panel) and coeval MAI (bottom panel) for three days during CR 2009. The dates
are indicated on the top of the panels.}
\end{figure}

Figures~2--3 display images of the frequency shifts averaged over all modes
(upper panels) as a function of the position of the solar disk for three days 
during CR 2009 and CR 2058, respectively.  The lower panels in each figure
correspond to the coeval MAI. For each of the images, each pixel corresponds
to a single tile in the dense-pack mosaic. 
Comparing the top and bottom panels, it is clearly seen that the tiles with
large frequency shifts match very well with the tiles at the same location
having a large magnetic field. 
Thus, areas of high frequency shifts  appear as active regions  and viceversa.
This confirms the findings of \citet{tripathy-2000SoPh..192..363H} that local
frequency shift acts as a tracer of magnetic activity.  However during the
low-activity period (Figure~\ref{tripathy-fig3}), where both the shifts and MAI
are small, we note a few tiles with positive frequency shifts that have  
no counterparts in the MAI
image implying a weaker agreement. In order to estimate how well we can
associate the locations of active regions as locations of frequency shifts, we
calculate the Pearson's correlation coefficient ($r_p$) between
the shifts and MAI for each of the three ring-days. These are found to be 
0.91, 0.93, and 0.88 and   0.74, 0.77, and 0.85, for CR 2009 and 2058,
respectively and confirms that the correlation between shifts and the surface
magnetic activity during the two activity periods are significantly different. 
This result is  consistent with the recent findings inferred from global
modes \citep{tripathy-jain09}. 

Thus the argument that the solar-cycle variations in the
global mode frequencies  are due to global averaging of the local effect of
active regions \citep{tripathy-2001ESASP.464..143H} is only partially supported
by our analysis.  We believe that the  weak component of the magnetic
field, e.g. ubiquitous horizontal field or turbulent field, must be taken into
account to fully explain the frequency shifts, particularly during the 
minimal-activity phase of the solar cycle. 

\begin{figure}  
  \centering
     \includegraphics[width=10cm]{\figspath/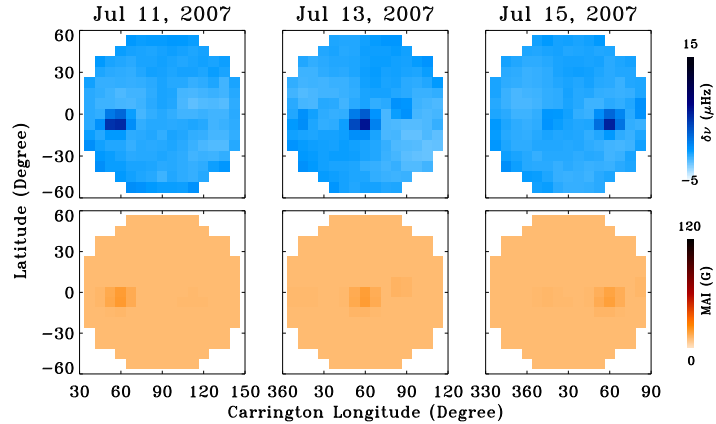}
    \caption[]{\label{tripathy-fig4}
Same as Figure~\ref{tripathy-fig3} but for low-activity period corresponding to
CR
2058. The dates are indicated on the top of the panels.}
\end{figure}

\begin{acknowledgement}
We thank John Leibacher for a critical reading of the manuscript. This research 
was supported in part by NASA grants NNG 05HL41I and NNG 08EI54I.
This work utilizes data obtained by the Global Oscillation Network
Group program, managed by the National Solar Observatory, which
is operated by AURA, Inc. under a cooperative agreement with the
National Science Foundation. The data were acquired by instruments
operated by the Big Bear Solar Observatory, High Altitude Observatory,
Learmonth Solar Observatory, Udaipur Solar Observatory, Instituto de
Astrof\'{\i}sica de Canarias, and Cerro Tololo Interamerican
Observatory. This work also utilizes 96-minute magnetograms from SOI/MDI on
 board {\em Solar and Heliospheric Observatory} (SOHO).  
SOHO is a  project of international cooperation between ESA and NASA. 
 
  \end{acknowledgement}

\begin{small}

\end{small}

\end{document}

%% file: rr-latexdefs.tex

\def\thisvolume{these proceedings}

\def\aj{{AJ}}			
\def\araa{{ARA\&A}}		
\def\apj{{ApJ}}			
\def\apjl{{ApJ}}		
\def\apjs{{ApJS}}		
\def\ao{{Appl.\ Optics}} 
\def\apss{{Ap\&SS}}		
\def\aap{{A\&A}}		
\def\aapr{{A\&A~Rev.}}		
\def\aaps{{A\&AS}}		
\def\an{{Astron.\ Nachrichten}}
\def\aspcs{{ASP Conf.\ Ser.}}
\def\azh{{AZh}}			
\def\baas{{BAAS}}		
\def\jrasc{{JRASC}}		
\def\memras{{MmRAS}}		
\def\mnras{{MNRAS}}
\def\nat{{Nat}}		
\def\pra{{Phys.\ Rev.\ A}} 
\def\prb{{Phys.\ Rev.\ B}}		
\def\prc{{Phys.\ Rev.\ C}}		
\def\prd{{Phys.\ Rev.\ D}}		
\def\prl{{Phys.\ Rev.\ Lett}}	
\def\pasp{{PASP}}
\def\pasj{{PASJ}}		
\def\qjras{{QJRAS}}
\def\science{{Sci}}		
\def\skytel{{S\&T}}		
\def\solphys{{Solar\ Phys.}} 
\def\sovast{{Soviet\ Ast.}}  
\def\ssr{{Space\ Sci.\ Rev.}}
\def\svassp{{Astrophys.\ Space Science Proc.}}
\def\zap{{ZAp}}			
\let\astap=\aap
\let\apjlett=\apjl
\let\apjsupp=\apjs

\def\ion#1#2{{\rm #1}\,{\uppercase{#2}}}  
\def\deg{\hbox{$^\circ$}}
\def\sun{\hbox{$\odot$}}
\def\earth{\hbox{$\oplus$}}
\def\la{\mathrel{\hbox{\rlap{\hbox{\lower4pt\hbox{$\sim$}}}\hbox{$<$}}}}
\def\ga{\mathrel{\hbox{\rlap{\hbox{\lower4pt\hbox{$\sim$}}}\hbox{$>$}}}}
\def\sq{\hbox{\rlap{$\sqcap$}$\sqcup$}}
\def\arcmin{\hbox{$^\prime$}}
\def\arcsec{\hbox{$^{\prime\prime}$}}
\def\fd{\hbox{$.\!\!^{\rm d}$}}
\def\fh{\hbox{$.\!\!^{\rm h}$}}
\def\fm{\hbox{$.\!\!^{\rm m}$}}
\def\fs{\hbox{$.\!\!^{\rm s}$}}
\def\fdg{\hbox{$.\!\!^\circ$}}
\def\farcm{\hbox{$.\mkern-4mu^\prime$}}
\def\farcs{\hbox{$.\!\!^{\prime\prime}$}}
\def\fp{\hbox{$.\!\!^{\scriptscriptstyle\rm p}$}}
\def\micron{\hbox{$\mu$m}}
\def\onehalf{\hbox{$\,^1\!/_2$}}	
\def\onethird{\hbox{$\,^1\!/_3$}}
\def\twothirds{\hbox{$\,^2\!/_3$}}
\def\onequarter{\hbox{$\,^1\!/_4$}}
\def\threequarters{\hbox{$\,^3\!/_4$}}
\def\ubv{\hbox{$U\!BV$}}		
\def\ubvr{\hbox{$U\!BV\!R$}}		
\def\ubvri{\hbox{$U\!BV\!RI$}}		
\def\ubvrij{\hbox{$U\!BV\!RI\!J$}}		
\def\ubvrijh{\hbox{$U\!BV\!RI\!J\!H$}}		
\def\ubvrijhk{\hbox{$U\!BV\!RI\!J\!H\!K$}}		
\def\ub{\hbox{$U\!-\!B$}}		
\def\bv{\hbox{$B\!-\!V$}}		
\def\vr{\hbox{$V\!-\!R$}}		
\def\ur{\hbox{$U\!-\!R$}}


\def\labelitemi{{\bf --}}  

\def\rmit#1{{\it #1}}              
\def\rmit#1{{\rm #1}}              
\def\etal{\rmit{et al.}}           
\def\etc{\rmit{etc.}}           
\def\ie{\rmit{i.e.,}}              
\def\eg{\rmit{e.g.,}}              
\def\cf{cf.}                       
\def\viz{\rmit{viz.}}
\def\vs{\rmit{vs.}}

\def\rot{\hbox{\rm rot}}
\def\div{\hbox{\rm div}}
\def\lesssim{\mathrel{\hbox{\rlap{\hbox{\lower4pt\hbox{$\sim$}}}\hbox{$<$}}}}
\def\gtrsim{\mathrel{\hbox{\rlap{\hbox{\lower4pt\hbox{$\sim$}}}\hbox{$>$}}}}
\def\dif{\: {\rm d}}                       
\def\ep{\:{\rm e}^}                        
\def\dash{\hbox{$\,-\,$}}                  
\def\is{\!=\!}                             

\def\starname#1#2{${#1}$\,{\rm {#2}}}  
\def\Teff{\hbox{$T_{\rm eff}$}}   

\def\kms{\hbox{km$\;$s$^{-1}$}}
\def\Mxcm{\hbox{Mx\,cm$^{-2}$}}    

\def\Bapp{\hbox{$B_{\rm app}$}}    

\def\komega{($k, \omega$)}                 
\def\kf{($k_h,f$)}                         
\def\VminI{\hbox{$V\!\!-\!\!I$}}           
\def\IminI{\hbox{$I\!\!-\!\!I$}}           
\def\VminV{\hbox{$V\!\!-\!\!V$}}           
\def\Xt{\hbox{$X\!\!-\!t$}}                

\def\level #1 #2#3#4{$#1 \: ^{#2} \mbox{#3} ^{#4}$}   

\def\specchar#1{\uppercase{#1}}    
\def\AlI{\mbox{Al\,\specchar{i}}}  
\def\BI{\mbox{B\,\specchar{i}}} 
\def\BII{\mbox{B\,\specchar{ii}}}  
\def\BaI{\mbox{Ba\,\specchar{i}}}  
\def\BaII{\mbox{Ba\,\specchar{ii}}} 
\def\CI{\mbox{C\,\specchar{i}}} 
\def\CII{\mbox{C\,\specchar{ii}}} 
\def\CIII{\mbox{C\,\specchar{iii}}} 
\def\CIV{\mbox{C\,\specchar{iv}}} 
\def\CaI{\mbox{Ca\,\specchar{i}}} 
\def\CaII{\mbox{Ca\,\specchar{ii}}} 
\def\CaIII{\mbox{Ca\,\specchar{iii}}} 
\def\CoI{\mbox{Co\,\specchar{i}}} 
\def\CrI{\mbox{Cr\,\specchar{i}}} 
\def\CriI{\mbox{Cr\,\specchar{ii}}} 
\def\CsI{\mbox{Cs\,\specchar{i}}} 
\def\CsII{\mbox{Cs\,\specchar{ii}}} 
\def\CuI{\mbox{Cu\,\specchar{i}}} 
\def\FeI{\mbox{Fe\,\specchar{i}}} 
\def\FeII{\mbox{Fe\,\specchar{ii}}} 
\def\FeIX{\mbox{Fe\,\specchar{ix}}}
\def\FeX{\mbox{Fe\,\specchar{x}}}
\def\FeXVI{\mbox{Fe\,\specchar{xvi}}}
\def\FrI{\mbox{Fr\,\specchar{i}}}
\def\HI{\mbox{H\,\specchar{i}}} 
\def\HII{\mbox{H\,\specchar{ii}}} 
\def\Hmin{\hbox{\rmH$^{^{_{\scriptstyle -}}}$}}      
\def\Hemin{\hbox{{\rm He}$^{^{_{\scriptstyle -}}}$}} 
\def\HeI{\mbox{He\,\specchar{i}}} 
\def\HeII{\mbox{He\,\specchar{ii}}} 
\def\HeIII{\mbox{He\,\specchar{iii}}} 
\def\KI{\mbox{K\,\specchar{i}}} 
\def\KII{\mbox{K\,\specchar{ii}}} 
\def\KIII{\mbox{K\,\specchar{iii}}} 
\def\LiI{\mbox{Li\,\specchar{i}}} 
\def\LiII{\mbox{Li\,\specchar{ii}}} 
\def\LiIII{\mbox{Li\,\specchar{iii}}} 
\def\MgI{\mbox{Mg\,\specchar{i}}} 
\def\MgII{\mbox{Mg\,\specchar{ii}}} 
\def\MgIII{\mbox{Mg\,\specchar{iii}}} 
\def\MnI{\mbox{Mn\,\specchar{i}}} 
\def\NI{\mbox{N\,\specchar{i}}}
\def\NaI{\mbox{Na\,\specchar{i}}}
\def\NaII{\mbox{Na\,\specchar{ii}}}
\def\NaIII{\mbox{Na\,\specchar{iii}}} 
\def\NiI{\mbox{Ni\,\specchar{i}}} 
\def\NiII{\mbox{Ni\,\specchar{ii}}}
\def\NiIII{\mbox{Ni\,\specchar{iii}}} 
\def\OI{\mbox{O\,\specchar{i}}} 
\def\OVI{\mbox{O\,\specchar{vi}}}
\def\RbI{\mbox{Rb\,\specchar{i}}} 
\def\SII{\mbox{S\,\specchar{ii}}} 
\def\SiI{\mbox{Si\,\specchar{i}}} 
\def\SiII{\mbox{Si\,\specchar{ii}}} 
\def\SrI{\mbox{Sr\,\specchar{i}}}
\def\SrII{\mbox{Sr\,\specchar{ii}}}
\def\TiI{\mbox{Ti\,\specchar{i}}} 
\def\TiII{\mbox{Ti\,\specchar{ii}}} 
\def\TiIII{\mbox{Ti\,\specchar{iii}}} 
\def\TiIV{\mbox{Ti\,\specchar{iv}}} 
\def\VI{\mbox{V\,\specchar{i}}} 
\def\HtwoO{\mbox{H$_2$O}}        
\def\Otwo{\mbox{O$_2$}}          

\def\Halpha{\mbox{H\hspace{0.1ex}$\alpha$}} 
\def\Ha{\mbox{H\hspace{0.2ex}$\alpha$}}
\def\Hbeta{\mbox{H\hspace{0.2ex}$\beta$}}
\def\Hgamma{\mbox{H\hspace{0.2ex}$\gamma$}}
\def\Hdelta{\mbox{H\hspace{0.2ex}$\delta$}}
\def\Hepsilon{\mbox{H\hspace{0.2ex}$\epsilon$}}
\def\Hzeta{\mbox{H\hspace{0.2ex}$\zeta$}}
\def\Lyalpha{\mbox{Ly$\hspace{0.2ex}\alpha$}}
\def\Lybeta{\mbox{Ly$\hspace{0.2ex}\beta$}}
\def\Lygamma{\mbox{Ly$\hspace{0.2ex}\gamma$}}
\def\Lycont{\mbox{Ly\hspace{0.2ex}{\small cont}}}
\def\Baalpha{\mbox{Ba$\hspace{0.2ex}\alpha$}}
\def\Babeta{\mbox{Ba$\hspace{0.2ex}\beta$}}
\def\Bacont{\mbox{Ba\hspace{0.2ex}{\small cont}}}
\def\Paalpha{\mbox{Pa$\hspace{0.2ex}\alpha$}}
\def\Bralpha{\mbox{Br$\hspace{0.2ex}\alpha$}}

\def\NaD{\mbox{Na\,\specchar{i}\,D}}    
\def\NaDone{\mbox{Na\,\specchar{i}\,\,D$_1$}}
\def\NaDtwo{\mbox{Na\,\specchar{i}\,\,D$_2$}}
\def\NaID{\mbox{Na\,\specchar{i}\,\,D}}
\def\NaIDone{\mbox{Na\,\specchar{i}\,\,D$_1$}}
\def\NaIDtwo{\mbox{Na\,\specchar{i}\,\,D$_2$}}
\def\Done{\mbox{D$_1$}}
\def\Dtwo{\mbox{D$_2$}}

\def\Mgbone{\mbox{Mg\,\specchar{i}\,b$_1$}}
\def\Mgbtwo{\mbox{Mg\,\specchar{i}\,b$_2$}}
\def\Mgbthree{\mbox{Mg\,\specchar{i}\,b$_3$}}
\def\MgIb{\mbox{Mg\,\specchar{i}\,b}}
\def\MgIbone{\mbox{Mg\,\specchar{i}\,b$_1$}}
\def\MgIbtwo{\mbox{Mg\,\specchar{i}\,b$_2$}}
\def\MgIbthree{\mbox{Mg\,\specchar{i}\,b$_3$}}

\def\CaIIK{\mbox{Ca\,\specchar{ii}\,K}}       
\def\CaIIH{\mbox{Ca\,\specchar{ii}\,H}}
\def\CaIIHK{\mbox{Ca\,\specchar{ii}\,H\,\&\,K}}
\def\HK{\mbox{H\,\&\,K}}
\def\Kthree{\mbox{K$_3$}}      
\def\Hthree{\mbox{H$_3$}}
\def\Ktwo{\mbox{K$_2$}}
\def\Htwo{\mbox{H$_2$}}
\def\Kone{\mbox{K$_1$}}     
\def\Hone{\mbox{H$_1$}}     
\def\KtwoV{\mbox{K$_{2V}$}}
\def\KtwoR{\mbox{K$_{2R}$}}
\def\KoneV{\mbox{K$_{1V}$}}
\def\KoneR{\mbox{K$_{1R}$}}
\def\HtwoV{\mbox{H$_{2V}$}}
\def\HtwoR{\mbox{H$_{2R}$}}
\def\HoneV{\mbox{H$_{1V}$}}
\def\HoneR{\mbox{H$_{1R}$}}

\def\hk{\mbox{h\,\&\,k}}
\def\kthree{\mbox{k$_3$}}    
\def\hthree{\mbox{h$_3$}}
\def\ktwo{\mbox{k$_2$}}
\def\htwo{\mbox{h$_2$}}
\def\kone{\mbox{k$_1$}}     
\def\hone{\mbox{h$_1$}}     
\def\ktwoV{\mbox{k$_{2V}$}}
\def\ktwoR{\mbox{k$_{2R}$}}
\def\koneV{\mbox{k$_{1V}$}}
\def\koneR{\mbox{k$_{1R}$}}
\def\htwoV{\mbox{h$_{2V}$}}
\def\htwoR{\mbox{h$_{2R}$}}
\def\honeV{\mbox{h$_{1V}$}}
\def\honeR{\mbox{h$_{1R}$}}